\documentclass[prd,amsmath,amssymb,showpacs,superscriptaddress,nofootinbib,twocolumn]{revtex4-1}
\usepackage{graphicx}
\usepackage{epsfig,graphics,subfigure,psfrag,amsmath,amssymb}
\usepackage{dcolumn}
\usepackage{bm}
\usepackage{overpic}
\usepackage{xspace}

\usepackage{multirow}
\usepackage{epstopdf}
\usepackage[colorlinks,linkcolor=blue,anchorcolor=blue,citecolor=blue]{hyperref}
\begin{document}
\title{\boldmath Search for the decay $\eta'\to\gamma\gamma\eta$}

\author{M.~Ablikim$^{1}$, M.~N.~Achasov$^{10,d}$, P.~Adlarson$^{59}$, S. ~Ahmed$^{15}$, M.~Albrecht$^{4}$, M.~Alekseev$^{58A,58C}$, A.~Amoroso$^{58A,58C}$, F.~F.~An$^{1}$, Q.~An$^{55,43}$, Y.~Bai$^{42}$
% ...
\\
\vspace{0.2cm}
(BESIII Collaboration)\\
\vspace{0.2cm} {\it
$^{1}$ Institute of High Energy Physics, Beijing 100049, People's Republic of China\\
% ...
$^{4}$ Bochum Ruhr-University, D-44780 Bochum, Germany\\
% ...
$^{10}$ G.I. Budker Institute of Nuclear Physics SB RAS (BINP), Novosibirsk 630090, Russia\\
% ...
$^{15}$ Helmholtz Institute Mainz, Johann-Joachim-Becher-Weg 45, D-55099 Mainz, Germany\\
% ...
$^{42}$ Southeast University, Nanjing 211100, People's Republic of China\\
$^{43}$ State Key Laboratory of Particle Detection and Electronics, Beijing 100049, Hefei 230026, People's Republic of China\\
% ...
$^{55}$ University of Science and Technology of China, Hefei 230026, People's Republic of China\\
% ...
$^{58}$ (A)University of Turin, I-10125, Turin, Italy; (B)University of Eastern Piedmont, I-15121, Alessandria, Italy; (C)INFN, I-10125, Turin, Italy\\
$^{59}$ Uppsala University, Box 516, SE-75120 Uppsala, Sweden\\
% ...
$^{d}$ Also at the Novosibirsk State University, Novosibirsk, 630090, Russia\\
% ...
}
}

\noaffiliation{}


\begin{abstract}
  Using a data sample of $1.31\times10^{9} ~J/\psi$ events collected with the BESIII detector, a search for $\eta'\to\gamma\gamma\eta$ via $J/\psi\to\gamma\eta'$ is performed for the first time. No significant $\eta'$ signal is observed in the $\gamma\gamma\eta$ invariant mass spectrum, and the 
  branching fraction of $\eta'\to\gamma\gamma\eta$ is determined to be less than $1.33 \times 10^{-4}$ at the 90$\%$ confidence level.
\end{abstract}

\maketitle

\section{INTRODUCTION}

The $\eta'$ meson provides a unique opportunity for understanding the distinct symmetry-breaking mechanisms present in low-energy Quantum Chromodynamics (QCD)~\cite{J.S1949}-\cite{A.K2009}, and its decays play an important role in exploring the effective theory of QCD at low energy~\cite{J.G1985}. Within the frameworks of the Linear $\sigma$ model and
the Vector Meson Dominance (VMD) model~\cite{R.J2010, R.E2012},  the branching fractions of
$\eta'\to\gamma\gamma\pi^{0}$ and $\eta'\to\gamma\gamma\eta$ are predicted to be
3.8 $\times 10^{-3}$ and 2.0 $\times 10^{-4}$~\cite{R.E2012}, respectively. The dominant contributions come from the vector meson exchange processes, where for $\eta'\to\gamma\gamma\pi^{0}$, the $\omega$ contributes 80.2$\%$ of the total VMD signal, while the $\rho$ contributes 4.6$\%$. For $\eta'\to\gamma\gamma\eta$, $\rho$ and $\omega$ contribute 59.9$\%$ and 15.8$\%$, respectively.

Recently using $1.31\times10^9$ $J/\psi$ events, BESIII reported the study of $\eta'\to\gamma\gamma\pi^{0}$ for the first time,
and the branching fraction of $\eta'\to\gamma\gamma\pi^{0}$ was determined to be $(32.0 \pm 0.7 \pm 2.3)\times 10^{-4}$~\cite{djp}. By excluding
the intermediate contributions from $\omega(\rho^0)\rightarrow\gamma\pi^0$, the so-called non-resonant branching fraction of $\eta'\to\gamma\gamma\pi^{0}$ was determined to be $(6.16 \pm 0.64\pm 0.67)\times$ 10$^{-4}$~\cite{djp}, which confirmed the theoretical prediction and indicated that this decay was dominated by the VMD processes.

Unlike $\eta'\to\gamma\gamma\pi^{0}$ decay, the $\eta^\prime\rightarrow\gamma\gamma\eta$ decay has not been observed to date. The most stringent upper limit, reported by GAMS-4$\pi$ setup, on the branching fraction of this decay is $8\times 10^{-4}$ at the 90\% confidence level (CL)~\cite{GAMS4pi}. The BESIII experiment using $J/\psi$ radiative decays has observed a series of $\eta^\prime$ new decay modes~\cite{eta'topipieta}-\cite{eta'to4pi}, and in this paper we present a search for $\eta'\to\gamma\gamma\eta$ in the $J/\psi$ radiative decay.

\section{DETECTOR AND MONTE CARLO SIMULATION}

The BESIII detector is a magnetic spectrometer~\cite{BES} located at the Beijing Electron Position Collider (BEPCII)~\cite{Yu:IPAC2016-TUYA01}. The cylindrical core of the BESIII detector consists of a helium-based multilayer drift chamber (MDC), a plastic scintillator time-of-flight system (TOF), and a CsI(Tl) electromagnetic calorimeter (EMC), which are all enclosed in a superconducting solenoidal magnet providing a 1.0~T (0.9~T in 2012) magnetic field. The solenoid is supported by an octagonal flux-return yoke with resistive plate counter muon identifier modules interleaved with steel. The acceptance of charged particles and photons is 93\% over $4\pi$ solid angle. The charged-particle momentum resolution at $1~{\rm GeV}/c$ is $0.5\%$, and the $dE/dx$ resolution is $6\%$ for the electrons from Bhabha scattering. The EMC measures photon energies with a resolution of $2.5\%$ ($5\%$) at $1$~GeV in the barrel (end cap) region. The time resolution of the TOF barrel part is 68~ps, while that of the end cap part is 110~ps.

Monte Carlo (MC) simulations are used to estimate backgrounds and determine the detection efficiencies. The {\sc geant4}-based~\cite{ref:geant4} simulation software {\sc boost}~\cite{ref:boost} includes the geometric and material description
of the BESIII detector, detector response, and digitization models, as well as the tracking of the detector running conditions and performance.  Production of the charmonium state $J/\psi$ is simulated with {\sc kkmc}~\cite{KKMC}, while the decays are generated with {\sc evtgen}~\cite{EVTGEN} for known decay modes with branching fractions taken from the Particle Data Group (PDG)~\cite{C.P2016} and by {\sc lundcharm}~\cite{ref:lundcharm} for the remaining unknown decays.

In this analysis, the program {\sc evtgen} is used to generate a $J/\psi\to\gamma\eta'$ MC sample with an angular distribution of $1 + \cos^2\theta_{\gamma}$, where $\theta_{\gamma}$ is the polar angle of the radiative photon in the $J/\psi$ rest frame.
  The decays $\eta'\to\gamma\omega$ ($\rho$), $\omega$ ($\rho$) $\to\gamma\eta$ are generated using the VMD model~\cite{R.J2010, R.E2012} with $\omega$ ($\rho$) exchange.  For the non-resonant $\eta'\to\gamma\gamma\eta$ decay, the VMD model is also used to generate the MC sample with  $\omega(1420)$ or $\rho(1450)$ exchange.
  We use a sample of $1.225 \times 10^{9}$ simulated $J/\psi$ events to study the backgrounds in which the $J/\psi$ decays generically (inclusive MC sample). The analysis is
performed in the framework of the BESIII offline software system~\cite{BOSS} which incorporates the detector calibration, event reconstruction, and data storage.

\section{EVENT SELECTION AND BACKGROUND ESTIMATION}

In the reconstruction of $J/\psi\to\gamma\eta'$ with $\eta'\to\gamma\gamma\eta$ and $\eta\to\gamma\gamma$, candidate events must have no charged particle and at least five photons. Charged particles are identified by tracks in the active region of the MDC, corresponding to $|\cos\theta|<0.93$, where $\theta$ is the polar angle of the charged track with respect to the beam direction. They are also required to pass within $\pm 10$ cm of the interaction point in the beam direction and 1 cm of the beam line in the plane perpendicular to the beam.
The photon candidate showers must have minimum energy of 25 MeV in the barrel region ($|\cos\theta| < 0.8$) or 50 MeV in the end cap region ($0.86 < |\cos\theta| < 0.92$). Showers in the region between the barrel and the end caps are poorly measured and excluded. A requirement of EMC cluster timing with respect to the most energetic photon ($-500$ ns $<$ T $<$ 500 ns) is used to suppress electronic noise and energy deposits unrelated to the event. To select $J/\psi\to\gamma\eta'$, $\eta'\to\gamma\gamma\eta$ ($\eta\to\gamma\gamma$) signal events, only the events with exactly five photon candidates are selected, and the most energetic photon is taken as the radiative photon from the $J/\psi$ decay.

A four-constraint (4C) kinematic fit imposing energy-momentum conservation is performed to the $\gamma\gamma\gamma\gamma\gamma$ hypothesis and the $\chi^2$ is required to be less than 200.
To distinguish the photons from $\eta^\prime$ and $\eta$ decays, a variable $\delta^{2}_{\eta'\eta}$, defined as $\delta^{2}_{\eta^\prime\eta} =
(\frac{M(\gamma\gamma\eta)-m(\eta')}{\sigma_{1}})^{2}+(\frac{M(\gamma\gamma)-m(\eta)}{\sigma_{2}})^{2}$, is introduced,
where $M(\gamma\gamma\eta)$ is the invariant mass of four of five selected photons (expect for the radiative photon) for reconstructing the $\eta'$ meson, $M(\gamma\gamma)$ is the invariant mass of photon pairs for reconstructing the $\eta$ meson, while $\sigma_1$ = 11.7 MeV/c$^2$ and $\sigma_2$ = 9.7 MeV/c$^2$ are the mass resolutions of $\eta'$ and $\eta$, respectively, obtained from the MC simulations, $m(\eta')$ and $m(\eta)$ are the $\eta'$ and $\eta$ nominal masses, respectively. We then require 
$|M(\gamma\gamma)-m(\eta)| < 50$ MeV/$c^{2}$ and the combination with the minimum value of $\delta^{2}_{\eta^\prime\eta}$ is chosen.
Next the $\delta^{2}_{\eta'\eta}$ is required to be less than $\delta^{2}_{\eta\eta}$, which is defined as $\delta^{2}_{\eta\eta}= (\frac{M_{1}(\gamma\gamma)-m(\eta)}{\sigma_{2}})^{2}+(\frac{M_{2}(\gamma\gamma)-m(\eta)}{\sigma_{2}})^{2}$, where $M_{1}(\gamma\gamma)$ and $M_{2}(\gamma\gamma)$ are the invariant masses of arbitrary two of five selected photons, to suppress the background events from $J/\psi\to\gamma\eta\eta$.

\begin{figure}[htbp]
\begin{center}
\mbox{
\begin{overpic}[width=0.4\textwidth]{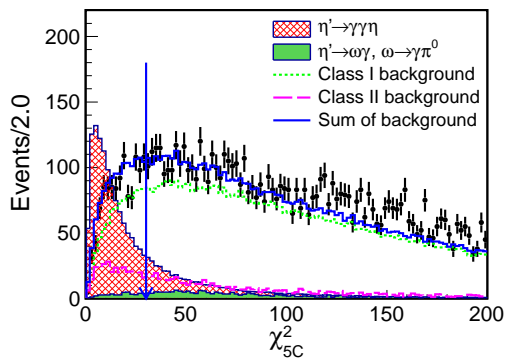}
\end{overpic}}
\caption{$\chi^2_{5C}$ distributions in MC simulations and data. Dots with error bars are data, the wide (blue) solid-curve is the sum of expected backgrounds from MC simulations, the grid area is from signal MC with arbitrary normalization, the (green) dotted-curve is the Class I ($J/\psi\to\gamma\eta'$, $\eta'\to\pi^{0}\pi^{0}\eta$) background, the (pink) dashed-curve is the Class II ($J/\psi\to\gamma\eta\pi^{0}$ and $J/\psi\to\omega\eta$ ($\omega\to\gamma\pi^{0}, \eta\to\gamma\gamma$)) background, and solid area is the peaking background.}
\end{center}
\label{chi5c}
\end{figure}

To improve the mass resolution and further suppress background events, a five-constraint (5C) kinematic fit imposing energy-momentum conservation with a $\eta$ mass constraint is performed under the $\gamma\gamma\gamma\eta$ hypothesis, where the $\eta$ candidate is reconstructed with the pair of photons described above, and the $\chi^2_{5C}$ is required to be less than 30. The $\chi^{2}_{5C}$ distribution is shown in Fig.~\ref{chi5c}. In addition, the invariant masses of all the two photon pairs are required not to be in the $\pi^0$ mass region, $|M(\gamma\gamma)-m(\pi^0)|>18$ MeV/c$^{2}$, to suppress the background events with $\pi^0$ in the final state.

To remove the mis-combinationed photon pairs in $\eta$ candidates, the $\eta$ decay angle $\theta_{\rm decay}$, defined as the polar angle of each photon in the corresponding $\gamma\gamma$ rest frame with respect to the $\eta$ direction in the $J/\psi$ rest frame, is required to satisfy $|\cos\theta_{\rm decay}| < 0.95$. An event is vetoed if any two of five selected photons (except for the combination for the $\eta$ candidate) satisfy $|M(\gamma\gamma)-m(\eta)| < 35$ MeV/$c^{2}$.

The resulting $\gamma\gamma\eta$ invariant mass distribution, after these requirements, is shown in Fig.~\ref{eta'fit}(a), where no significant $\eta^\prime$ peak is observed. Detailed MC studies indicate that the background events accumulating near the lower side of the $\eta'$ signal region are mainly from $J/\psi\to\gamma\eta'$, $\eta'\to\pi^{0}\pi^{0}\eta$ (Class I), which is shown as the dotted (green) curve in Fig.~\ref{eta'fit}(a). The peaking background is from $J/\psi\to\gamma\eta', \eta'\to\gamma\omega, \omega\to\gamma\pi^0$, and is shown as the solid area in Fig.~\ref{eta'fit}(a). The remaining background events are dominated by those $J/\psi$ decays without $\eta^\prime$ in the final states (Class II), e.g., $J/\psi\to\gamma\eta\pi^{0}$ and $J/\psi\to\omega\eta$ ($\omega\to\gamma\pi^{0}, \eta\to\gamma\gamma$) decays. They constitute a smooth distribution in the $\eta'$ signal region as illustrated by the dashed (pink) curve in Fig.~\ref{eta'fit}(a).

\section{SIGNAL YIELD AND BRANCHING FRACTION}
\begin{figure}[htbp]
\begin{center}
\mbox{
\begin{overpic}[width=0.4\textwidth]{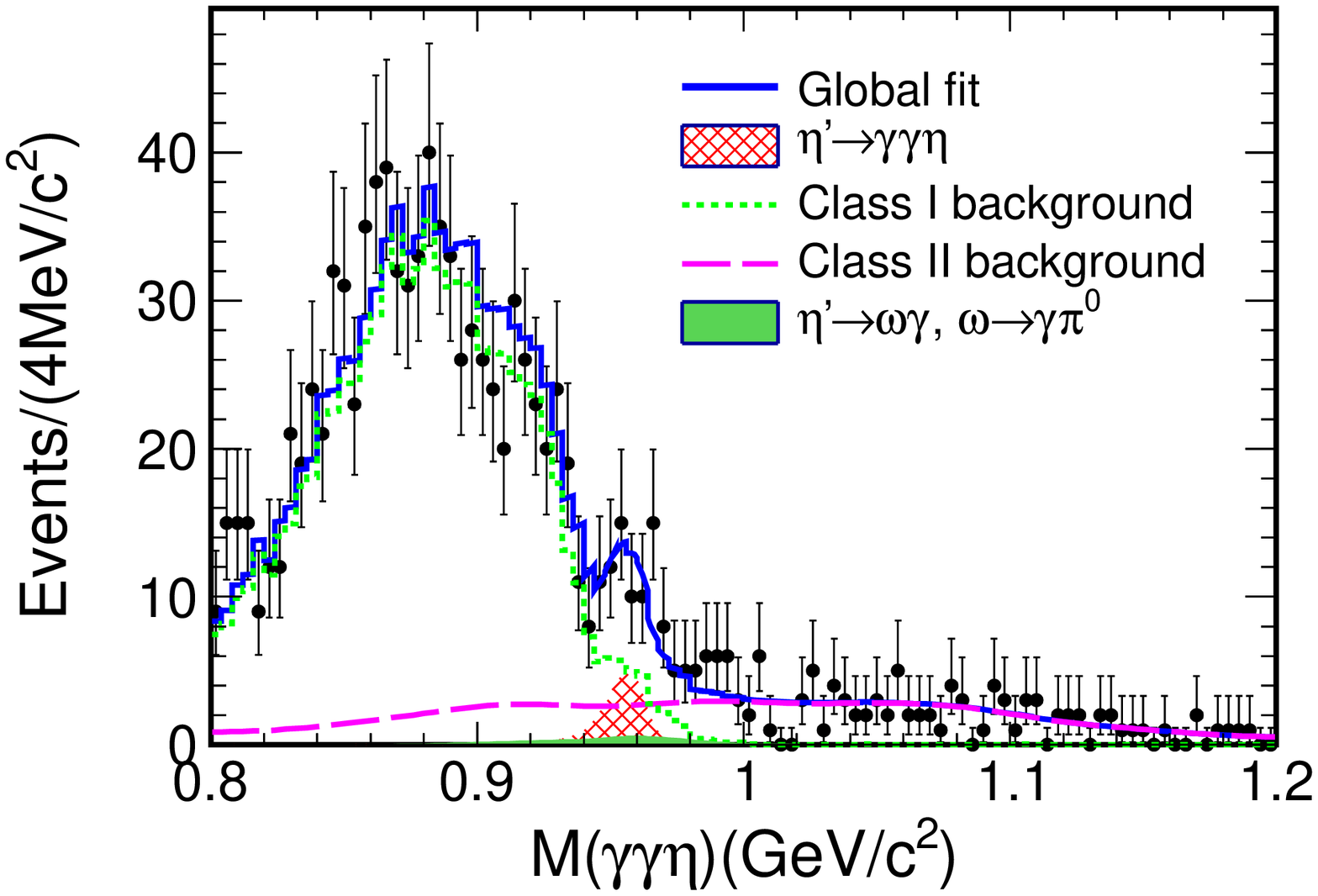}
\put(85,60){\color{black}  (a)}
\end{overpic}}
\mbox{
\begin{overpic}[width=0.4\textwidth]{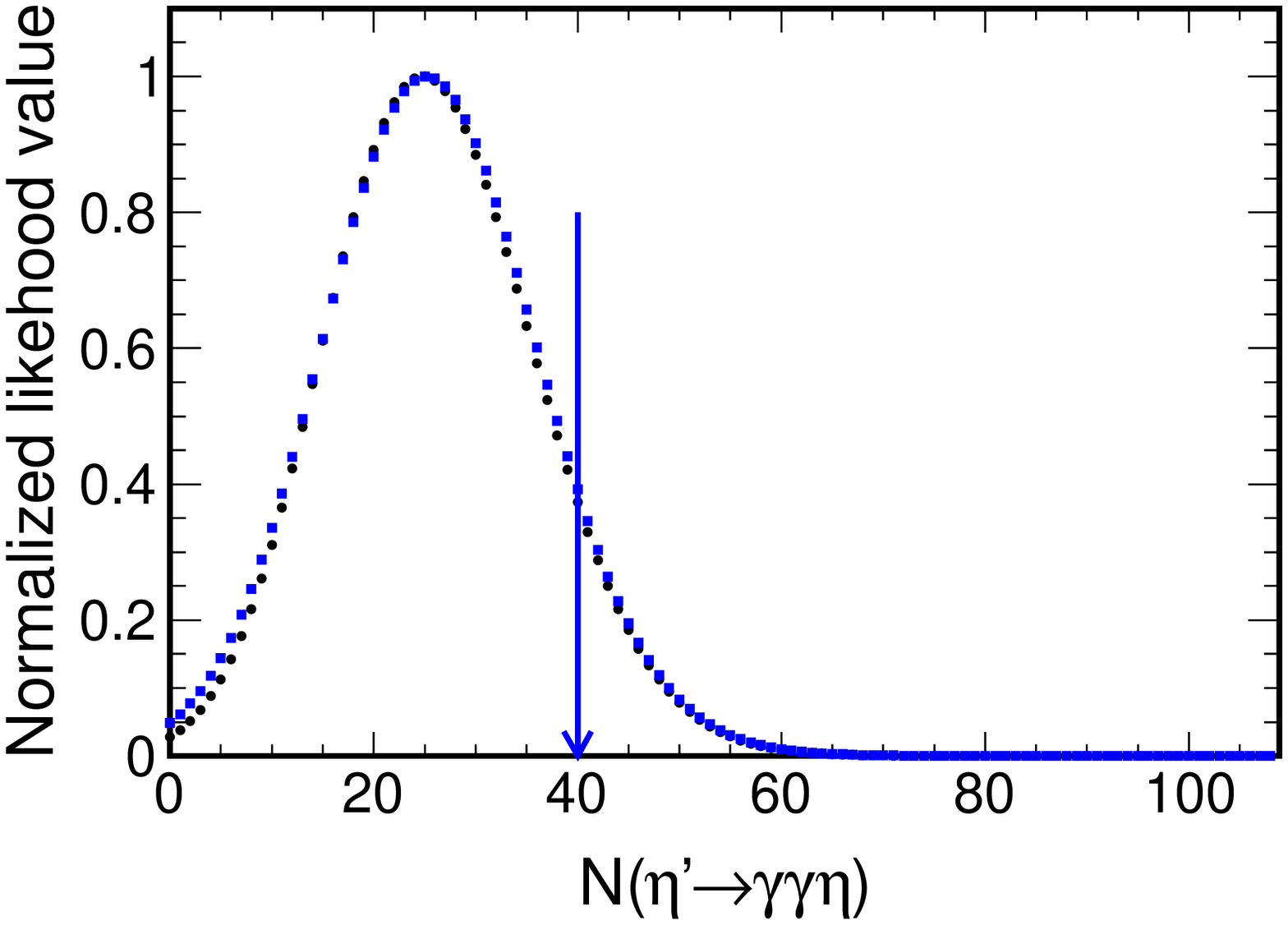}
\put(82,60){\color{black}  (b)}
\end{overpic}}
\caption{(a) Results of the fit to $M(\gamma\gamma\eta)$. The black dots with error bars are data, and the others are the results of the fit described in the text. 
(b) Likelihood distribution before (black
dots) and after (blue squares) taking into account systematic uncertainties (see Eq. (2)). The arrow is the position of the upper limit on the signal yields at 90$\%$ CL.}
\end{center}
\label{eta'fit}
\end{figure}

 An unbinned maximum likelihood fit to the $M(\gamma\gamma\eta)$ distribution is performed to determine the $\eta^\prime\rightarrow\gamma\gamma\eta$ signal yield. In the fit, the probability density function (PDF) for the signal component is represented by the signal MC shape, which is obtained from the signal MC sample generated with an incoherent mixture of $\rho, \omega$ and the non-resonant components according to the fractions from the theoretical prediction~\cite{R.J2010, R.E2012}. The Class I and Class II background shapes are obtained from MC simulations and fixed, but the numbers are free parameters. Both the shape and the yield for the peaking background are fixed to the MC simulation and their expected intensities. 
The fit shown in Fig.~\ref{eta'fit}(a) yields $24.9 \pm 10.3$ $\eta'\to\gamma\gamma\eta$ events with a statistical significance of 2.6$\sigma$, and the branching fraction is calculated from
\begin{equation}
B(\eta'\to\gamma\gamma\eta)= \frac{N_{obs}}{N_{J/\psi} \cdot \varepsilon \cdot B(\eta\to\gamma\gamma) \cdot B(J/\psi\to\gamma\eta')},
 \end{equation}
where $N_{obs}$ is the number of observed events determined from the fit to the $\gamma\gamma\eta$ mass spectrum,
$\varepsilon$ is the MC-determined detection efficiency, which is obtained from the signal MC sample described above;  $B(\eta\to\gamma\gamma)$ and $B(J/\psi\to\gamma\eta')$ are the branching fractions of $\eta\to\gamma\gamma$ and $J/\psi\to\gamma\eta'$ quoted from the PDG~\cite{C.P2016}, respectively.

With the number of signal events and a detection efficiency of 11.4\%
 the branching fraction is measured to be
 \begin{center}
 $B(\eta'\to\gamma\gamma\eta) = (8.25 \pm 3.41 \pm 0.72)\times 10^{-5}$,\\
 \end{center}
where the first uncertainty is statistical and the second systematic.

\section{SYSTEMATIC UNCERTAINTIES}

The systematic uncertainties on the upper limit measurement are summarized in Table~\ref{sys.}. The uncertainty due to the photon reconstruction is determined to be $0.5\%$ per photon in the EMC barrel and $1.5\%$ per photon in the EMC endcap~\cite{M.A2010}. Thus the uncertainty associated with the five reconstructed photons is $3\%$ ($0.6\%$ per photon) by weighting the uncertainties according to the polar angle distribution of the five photons from data. The uncertainties associated with the other selection criteria, e.g., kinematic fit with $\chi_{5C}^{2}<30$, the number of photons equal to 5, $\pi^0$ veto ($|M(\gamma\gamma)-m(\pi^{0})|>18 {\rm MeV/c^{2}}$) and $\cos\theta_{\rm decay}$,  are studied using the
$J/\psi\to\gamma\eta'\to\gamma\gamma\omega$, $\omega\to\gamma\pi^{0}$ decay control sample~\cite{djp}. The systematic uncertainty for each of the applied selection criteria is numerically estimated from the difference of the number of events with and without the corresponding requirement. The resultant efficiency differences between data and MC simulations ($2.7\%$, $0.5\%$, $1.9\%$, and 0.3\%, respectively) are taken as the corresponding systematic uncertainties.

To suppress the multi-$\eta$ backgrounds and remove the mis-combinations of photons, an event is vetoed if any two of five selected photons (except for the combination for the $\eta$ candidate) satisfy $|M(\gamma\gamma)-m(\eta)| < 35$ MeV/$c^{2}$. To estimate the systematic uncertainty, this requirement varied by $\pm$ 10 MeV/$c^{2}$, and the maximum change to the nominal result is taken as the systematic uncertainty.

The signal shape is obtained from the MC simulation in the nominal fit for the $\eta'$ decay. The uncertainty due to the signal shape is considered by convolving a Gaussian function to account for the difference in the mass resolution between data and MC simulation. In the fit to the $\gamma\gamma\eta$ distribution, the signal PDF is the signal MC shape convolving a Gaussian function with a fixed width of 1.5 MeV~\cite{djp}, and the changes of the signal yields is taken as the uncertainty due to the signal shape.

The uncertainty due to the Class I background and the peaking background are estimated by varying the numbers of expected background events by one standard deviation according to the uncertainties on the branching fractions values in PDG~\cite{C.P2016}.

To take into account the systematic uncertainty due to signal model (VMD model), fits with alternative signal models for the different components, for example, a coherent sum for the $\rho$, $\omega-$components and a uniform angular distribution in phase space for the non-resonant process is performed. The resultant changes in the branching fractions (involving efficiency changes) are taken as the uncertainty related to the signal model.

To take into account the systematic uncertainties associated with the fit of the mass spectrum coming from the background events and the fit range, alternative fits with different fit ranges, background shapes and the number of background events are performed. The largest number of the signal yield among these cases is chosen to calculate the upper limit of the branching fraction at the 90$\%$ CL.

The number of $J/\psi$ events is $N_{J/\psi} = (1310.6 \pm 7.0) \times 10^6$~\cite{M.A2017}, corresponding to an uncertainty of 0.5$\%$. The branching fractions for the $J/\psi\to\gamma\eta'$ and $\eta\to\gamma\gamma$ decays are taken from the PDG~\cite{C.P2016}, and the corresponding uncertainties are taken as a systematic uncertainty.

Assuming all systematic uncertainties in Table~\ref{sys.} are independent, the total systematic uncertainty, obtained from their quadratic sum, is $8.7\%$.

\begin{table}[htbp]
 \centering
 {\caption{Summary of relative systematic uncertainties for the upper limit on the branching fraction measurement (in $\%$).}
 \label{sys.}}
 \begin{tabular}{c|c}
 \hline
 \hline
 Source & Systematic uncertainties \\
 \hline
  Photon detection & 3.0 \\
  Kinematic fit (5C) & 2.7 \\
  Number of photons ($N_{\gamma} = 5$) & 0.5 \\
  $\pi^0$ and $\gamma$ veto & 1.9 \\
  $\cos\theta_{\rm decay}$ & 0.3 \\
  $\eta$ veto & 4.3 \\
  Signal shape & 3.2 \\
  Class I background  & 3.1 \\
  Peaking background & 0.8 \\
  Signal model & 2.9 \\
  Cited branching fractions & 3.3 \\
  Number of $J/\psi$ events & 0.5 \\
 \hline
 Total & 8.8 \\
 \hline
 \hline
 \end{tabular}
 \end{table}

\section{\boldmath $\eta'\to\gamma\gamma\eta$ upper Limit RESULTS}

Since no significant $\eta^\prime$ peak is seen, we use the Bayesian method to obtain the signal upper limit.
Unbinned maximum likelihood fits are performed on the $\gamma\gamma\eta$ mass spectrum with a series of input signal events, and the distribution of normalized likelihood values is taken as the PDF for the expected number of events.

The final upper limit on the branching fraction is determined by convolving the normalized likelihood curve $L(N)$ with the systematic uncertainties as a Gaussian function ($G(\mu, \sigma)=G(0, \sigma_{\rm sys})$) to obtain the smeared likelihood $L'(N')$, which is written as
\begin{equation}
L'(N')=\int\nolimits_0^{\infty} L(N)\dfrac{1}{\sqrt{2\pi}\sigma_{sys}}\exp[\dfrac{-(N'-N)^2}{2\sigma^2_{sys}}]dN,
\end{equation}
where $\sigma_{\rm sys}=N\cdot\sigma_{\rm rel}$, $N$ and $\sigma_{\rm rel}$ are the input signal yield and the corresponding uncertainty, respectively.
Figure~\ref{eta'fit}(b) shows the likelihood distribution before and after convolving the Gaussian function.  The upper limit on the number of $\eta^\prime\rightarrow\gamma\gamma\eta$ events, $N'_{UL}$, is determined to be 40 at the 90\% CL .
The corresponding upper limit on the branching fraction of $\eta'\to\gamma\gamma\eta$ is determined to be
$B(\eta'\to\gamma\gamma\eta)<1.33 \times 10^{-4}$
at the 90\% CL.

\section{SUMMARY}
With a data sample of 1.31 $\times 10^{9}$ $J/\psi$ events collected with the BESIII
detector, we report on a search for the doubly radiative decay $\eta'\to\gamma\gamma\eta$ for the first time, where the $\eta'$ meson is produced via the $J/\psi\to\gamma\eta'$ process. The observed signal yields in the $\gamma\gamma\eta$ invariant mass spectrum corresponds to 2.6$\sigma$, this signal corresponds to a branching fraction of $(8.25 \pm 3.41 \pm 0.72)\times 10^{-5}$. We also present an upper limit of the branching fraction of $1.33 \times 10^{-4}$ at the 90\% CL. The obtained result is in tension with a recent theoretical prediction of $2.0\times 10^{-4}$~\cite{R.E2012} within the frame work of the linear $\sigma$ model and the VMD model.

\color{black}
\begin{acknowledgments}
 The BESIII collaboration thanks the staff of BEPCII and the IHEP computing center for their strong support. This work is supported in part by National Key Basic Research Program of China under Contract No. 2015CB856700; National Natural Science Foundation of China (NSFC) under Contracts Nos. 11625523, 11635010, 11675184, 11735014, 11835012; the Chinese Academy of Sciences (CAS) Large-Scale Scientific Facility Program; Joint Large-Scale Scientific Facility Funds of the NSFC and CAS under Contracts Nos. U1532257, U1532258, U1732263, U1832207; CAS Key Research Program of Frontier Sciences under Contracts Nos. QYZDJ-SSW-SLH003, QYZDJ-SSW-SLH040; 100 Talents Program of CAS; INPAC and Shanghai Key Laboratory for Particle Physics and Cosmology; German Research Foundation DFG under Contracts Nos. Collaborative Research Center CRC 1044, FOR 2359; Istituto Nazionale di Fisica Nucleare, Italy; Koninklijke Nederlandse Akademie van Wetenschappen (KNAW) under Contract No. 530-4CDP03; Ministry of Development of Turkey under Contract No. DPT2006K-120470; National Science and Technology fund; The Knut and Alice Wallenberg Foundation (Sweden) under Contract No. 2016.0157; The Royal Society, UK under Contract No. DH160214; The Swedish Research Council; U. S. Department of Energy under Contracts Nos. DE-FG02-05ER41374, DE-SC-0010118, DE-SC-0012069; University of Groningen (RuG) and the Helmholtzzentrum fuer Schwerionenforschung GmbH (GSI), Darmstadt.
\end{acknowledgments}

\end{document}